\documentclass[]{article}
\usepackage{cite}
\usepackage{amsmath,amssymb,amsfonts}
\usepackage{graphicx}
\usepackage{textcomp}
\usepackage{color}
\usepackage{float}
\usepackage{bm}

\usepackage{algorithm}
\usepackage{algpseudocode}
\usepackage{amsmath}
\usepackage{graphics}
\usepackage{epsfig}
\usepackage{geometry}

\title{PET Image Reconstruction with Multiple Kernels and Multiple Kernel Space Regularizers}

\begin{document}

\date{}
	
	\author{Shiyao Guo \thanks{The Engineering Research Center of Metallurgical Automation and Measurement Technology,
			Wuhan University of Science and Technology, Wuhan, China. e-mails: wustguosy@foxmail.com; shengyuxia@wust.edu.cn; chaili@wust.edu.cn.},
		 Yuxia Sheng$^*$, 
		 Shenpeng Li\thanks{Health and Biosecurity, the Australian e-Health Research Centre, CSIRO, Parkville, VIC 3052, Australia. e-mail: shenpeng.li@csiro.au.}, Li Chai$^*$, 
		 Jingxin Zhang\thanks{The School of Software and Electrical Engineering, Swinburne University of Technology, Melbourne, Vic 3126, Australia. e-mail: jingxinzhang@swin.edu.au.}}
		
\maketitle

\begin{abstract}
Kernelized maximum-likelihood (ML) expectation maximization (EM) methods have recently gained prominence in PET image reconstruction, outperforming many previous state-of-the-art methods. But they are not immune to the problems of non-kernelized MLEM methods in potentially large reconstruction error and high sensitivity to iteration number. This paper demonstrates these problems by theoretical reasoning and experiment results, and provides a novel solution to solve these problems. The solution is a regularized kernelized MLEM with multiple kernel matrices and multiple kernel space regularizers that can be tailored for different applications. To reduce the reconstruction error and the sensitivity to iteration number, we present a general class of multi-kernel matrices and two regularizers consisting of kernel image dictionary and kernel image Laplacian quatradic, and use them to derive the single-kernel regularized EM and multi-kernel regularized EM algorithms for PET image reconstruction. These new algorithms are derived using the technical tools of multi-kernel combination in machine learning, image dictionary learning in sparse coding, and graph Laplcian quadratic in graph signal processing.
Extensive tests and comparisons on the simulated and in vivo data are presented to validate and evaluate the new algorithms, and demonstrate their superior performance and advantages over the kernelized MLEM and other conventional methods.

\end{abstract}


\section{Introduction}
\label{sec:introduction}

The maximum-likelihood (ML) expectation maximization (EM) and the
maximum a posteriori (MAP) expectation maximization are two popular classes of
methods for PET image reconstruction \cite{wernick2004}. They both use EM to iteratively approximate ML or MAP probability.

\begin{figure}[!t]
	\centerline{\includegraphics[width=1\columnwidth]{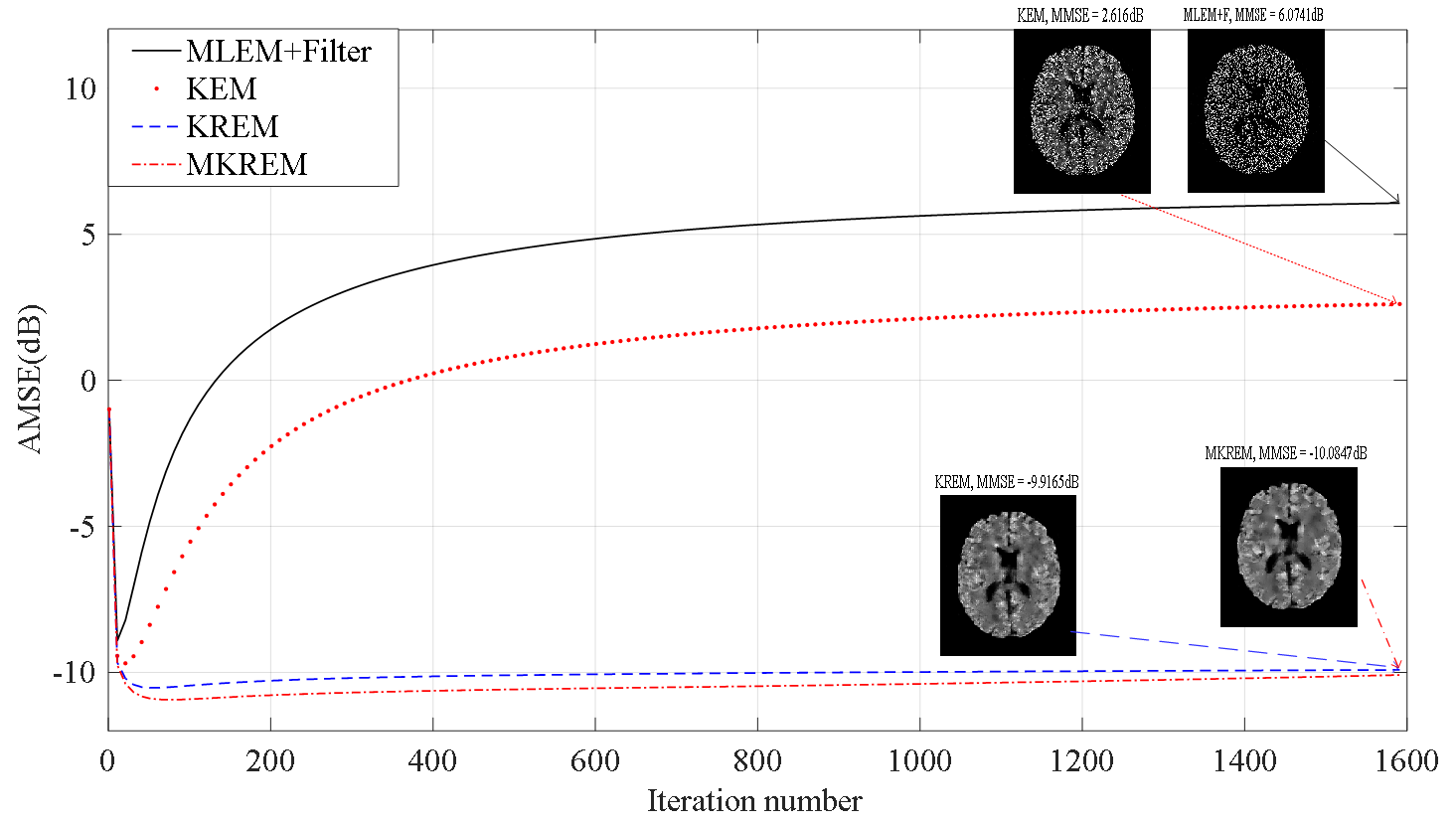}}
	\caption{Average mean square error (AMSE) of reconstruction vs number of iterations for MLEM, KEM and the proposed KREM and MKREM.}
	\label{fig1}
\end{figure}

As exemplified in Fig. \ref{fig1} (solid curve), the MLEM method has two main problems: its reconstruction error may decrease and then increases rapidly with iterations first, then slowly converges to very noisy images due to ML's inherent nature of large variance under noise \cite{wernick2004,barrett1999}. These problems become more severe in the low count situation. Hence, MLEM's iterated estimates must be stopped well before convergence and filtered for useable images. To overcome these problems, the MAPEM method was introduced in PET image reconstruction \cite{wernick2004,levitan1987,green1990,lalush1992}. This method is equivalent to a regularized MLEM estimation with the regularization functions (regularizers) introduced as the priors of PET images \cite{levitan1987}. Thanks to regularization, MAPEM's reconstruction error is much lower and its iterated estimates tend to converge much faster than its ML counterpart.

Because of its good numerical properties and convenience of incorporating
priors, the MAPEM method, also known as Bayesian method, has been
exploited in PET image reconstruction to incorporate various prior
information, especially the anatomical priors from multi-modality
imaging scanners, eg magnetic resonance (MR) image priors in PET-MR.
This has led to numerous reconstruction algorithms with
significant improvement of PET image quality and numerical robustness for
various applications of different imaging conditions, see
\cite{gindi1993,bowsher2004,ehrhardt2016,vunckx2012,jiao2014,tang2009,yang2018,tang2014} for examples.

Besides the MAPEM method, there are also other effective methods to introduce the priors in PET image reconstruction,
eg the methods of joint and synergistic reconstruction of PET and MR images \cite{sudarshan2018,mehranian2017,knoll2016}.
One of these methods is the kernelized expectation maximization (KEM) method recently introduced in \cite{hutchcroft2016,wang2015}. This method uses the prior data, such as MR anatomical images, to construct the kernel matrix and uses MLEM in the kernel space to iteratively estimate the kernel coefficients of the image, which in turn are mapped by the kernel
matrix to the reconstructed image. It has shown superior performance in enhancing the
spatial and temporal resolution and quantification in static and
dynamic PET imaging, with great potential in low dose PET imaging.
Following these promising results, a number of kernelized methods
have recently been presented for various PET image reconstruction
problems, see eg
\cite{bland2018,novosad2016,gong2018,bland20181,ellis2016,baikejiang2017,spencer2017}.

Despite the advantages of kernelized methods, they may suffer from the same problems of MLEM method. Fig. \ref{fig1} (dotted curve) shows an example of high-count static PET image reconstruction using KEM, and similar result is also given in \cite{wang2019}. It can be seen that KEM behaves similarly to MLEM and can be highly sensitive to the iteration number around its minimum reconstruction error. This can be even worse in low-count situation. 
Such sensitivity may severely degrade PET image
quality, since it is generally difficult in practice to preset the maximum
iteration number to the optimal value with the minimum reconstruction
error, see Section \ref{Ex-results} for examples.

The reason for kernelized methods' sensitivity to iteration number is the lack of
effective regularization in their kernel space ML estimation. From the kernel and regularization theory \cite{smola2003}, an optimization regularizer is generally a function of the optimizers (optimization variables) that varies with the optimizers, eg grows as the optimizers grow, so as to constrain the optimizers, and a kernel can function as a regularizer if it has the same property. This is the case in the MAPEM based  reconstruction methods, eg \cite{wernick2004,levitan1987,green1990,lalush1992}, where the regularizers are the functions of the estimated variables. Hence, they generally behave like the dashed and dashed-dotted curves in Fig. 1, with much lower reconstruction error and sensitivity to iteration number. But this is not the case in the kernelized methods, where the only possible regularizer, the kernel matrix, is constructed using prior data and does not vary with the estimated variables. Therefore, they generally lack effective regularization even though their kernel may be correlated with the estimated variables to some extent. For dynamic PET imaging, \cite{wang2019} has shown by simulations that KEM's sensitivity can be reduced by combining the spatial kernel with a temporal kernel. However, without supporting mathematical theory, the underlying mechanism and generality of this KEM variant are unclear. Moreover, this variant is inapplicable to static PET imaging where no temporal kernel is available. All these strongly suggest that some explicit and effective regularizations are needed in the kernelized methods in order to reduce their reconstruction error and sensitivity to iteration number.

The kernelized method was initially proposed to use only a single spatial kernel, and was later extended to include another temporal kernel or spatial kernel from other source for  improved performance \cite{wang2019, deidda2019}. However, these extensions are application specific and have not provided a general guidance for finding suitable kernels for different applications.

To address above problems, we propose in this work a novel framework for the kernelized PET image reconstruction. This framework is a general reformulation of kernel space MAPEM estimation with multi-kernel matrices and multiple kernel space regularizers that can be tailored to application requirements. To significantly reduce the reconstruction error and sensitivity to iteration number, we present a general class of multi-kernel matrices and two specific kernel space regularizers, and use them to derive the single-kernel regularized EM (KREM) and multi-kernel regularized EM (MKREM) algorithms for PET image reconstruction. These algorithms are derived using the technical tools of multiple kernel combination in machine learning \cite{taylor2004,cortes2009,onen2011}, image dictionary learning \cite{elad2010} in sparse coding, and graph Laplcian quadratic \cite{shuman 2013} in graph signal processing.
We further use simulated and in vivo data to validate and evaluate the performance of KREM and MKREM, and demonstrate their superior performance and advantages over the existing methods by comprehensive comparisons.

This paper is organized as follows.  Section \ref{Sect 2} presents our proposed method in detail. Section \ref{Ex-results} presents the simulation studies and comparisons of the proposed method with the existing methods. Section \ref{invivo} presents the in vivo data testing results of the proposed method and compares with those of the existing methods. Discussions and Conclusions are presented in Sections \ref{discus} and \ref{conclu}, respectively.

\section{Theory and Method}
\label{Sect 2}
\subsection{PET Image Reconstruction with Multi-regularizers }
Let $\bm p= [p_{1}, \cdots, p_i, \cdots, p_{N}]^{T}$ be the PET projection data vector with its elements being Poisson random variables having the log-likelihood function \cite{wang2015}
\begin{equation}
\label{LL}
L(\bm{p} | \bm{x})=\sum_{i=1}^{N} p_{i} \log \bar{p}_{i}-\bar{ p}_{i}-\log p_{i} !
\end{equation}
where $N$ is the total number of lines of response and $\bar{p}_{i}$ is the expectation of $p_i$.
The expectation of $\bm{p}$, $\bar{\bm{p}} = [\bar{p}_{1}, \cdots, \bar{p}_{i},\cdots, \bar{p}_{N}]^{T}$, is related to the unknown emission image $\bm x = [x_{1}, \cdots, x_{i}, \cdots, x_{M}]^T$ by the forward projection
\begin{equation}
\label{FPROJ}
\bar{\bm p} = \bm{H}\bm x + \bm{r}
\end{equation}
where  $x_i$ is the $i$th voxel intensity of the image $\bm x$, $M$ is the total number of voxels,
$\bm H \in {\cal R}^{N \times M}$ is the PET system matrix, and $\bm{r} = [r_{1},  \cdots, r_{i}, \cdots, r_{N}]^T $ is the expectation of random and scattered events.

An estimate of the unknown image $\bm{x}$ can be obtained by the maximum a posteriori (MAP) estimation of $\bm x$ \cite{green1990,wernick2004}
\begin{equation}
\label{MAP}
\hat{\bm{x}}=\underset{\bm{x} \geq 0}{\arg \max } \,\, P(\bm{p} | \bm{x}) P(\bm{x})/P(\bm{p}).
\end{equation}
Here $P(\bm p)$ is a normalizing constant independent of $\bm x$; $P(\bm{x})$ represents the prior knowledge of unknown image $\bm{x}$ and is assumed to have the Gibbs distribution \cite{oliver2013}
\begin{equation}
\label{PK}
P(\bm{x}) = e^{-V(\bm{x})}/W
\end{equation}
where $W$ is a normalizing constant and $V(x)$ the energy function. In this paper, we consider $V(\bm x)$ as a weighted sum of multiple prior functions $v_{i}(\bm{x})$
\begin{equation}
\label{MP}
V(\bm{x}) = \sum_{i = 1}^{m}\beta_{i}v_{i}(\bm{x})
\end{equation}
with the weighting parameters $\beta_{i}$.

Solving $\hat{\bm{x}}$ from (\ref{MAP}) is equivalent to maximizing the log-posterior probability with
\begin{equation}
\label{LMAP}
\hat{\bm{x}}=\underset{\bm{x} \geq 0}{\arg \max} \,\, (L(\bm{p} | \bm{x})- \sum_{i = 1}^{m}\beta_{i}v_{i}(\bm{x}))
\end{equation}
where $L(\bm{p} | \bm{x})$ is as defined in (\ref{LL}), and the contributions of $P(\bm p)$ 
and $W$ have been excluded since they are independent of $\bm x$. When all $\beta_i = 0$, (\ref{LMAP}) reduces to the ML estimation \cite{wang2015}. The multiple prior functions $\beta_{i}v_{i}(\bm{x})$ in $V(\bm x)$ introduce multiple regularizers in the ML estimation. 

The solution of (\ref{LMAP}) can be readily found by the regularized MLEM algorithm \cite{jiao2014,green1990,tang2014}
\begin{equation}
\label{RMAP}
\begin{aligned}
\bm x^{n+1}=(\bm x^{n} \circ \bm{H}^{T} (\bm p \oslash (\bm H \bm x^{n}+\bm r))) \oslash \\
(\bm{H}^{T} \bm{1}_{N}+ \sum_{i = 1}^{m}\beta_{i}\frac{\partial v_{i}(\bm x^{n})}{\partial \bm x^{n}}),
\end{aligned}
\end{equation}
where $\partial v_{i}(\bm x^{n})/\partial \bm x^{n} := [\partial v_{i}({\bm x})/\partial {\bm x}]_{{\bm x} = \bm x^{n}}$,
$\bm{x}^n$ is the estimate of $\bm x$ at the $n$th iteration, $\bm{1}_N \in R^N$ is an all 1 vector,  $\circ$ and $\oslash$ are Hadamard product and division, defined as $A \circ B = [a(i,j)b(i,j)]$ and $A \oslash B = [a(i,j)/b(i,j) ]$ for the matrices $A = [a(i,j) ]$ and $B = [b(i,j) ]$ of the same dimension.

The algorithm (\ref{RMAP}) includes the MLEM algorithm as a special case when all $\beta_i = 0$. The summation term of $(\beta_i \partial v_{i}(\bm x^{n})/\partial \bm x^{n})$'s in (\ref{RMAP}) regularizes the MLEM algorithm, which greatly accelerates convergence and reduces the variance at convergence \cite{wernick2004,green1990} as discussed before.

\subsection{ PET Image Reconstruction with Single-kernel and Multi-Regularizers}
Using the kernel theory \cite{taylor2004}, the unknown emission image $\bm x$  can be presented in the kernel space as \cite{wang2015}
\begin{equation}
\label{KIM}
\bm{x} = \bm K\bm a 
\end{equation}
where $\bm K \in {\cal R}^{M \times M}$ is a kernel (basis) matrix and ${\bm a} \in {\cal R}^M$ is the coefficient vector of $\bm x$ under $\bm K$.

The $ij$th element of 
$\bm K$ is given by
$K(i,j)= k( \bm{\tilde{x}}_{i},  \bm{\tilde{x}}_{j}) = \left \langle \phi (\bm{\tilde{x}}_{i}), \phi( \bm{\tilde{x}}_{j})\right\rangle, i, j \in \{1, 2, \cdots, M \}$, where $k( \bm{\tilde{x}}_{i}, \bm{\tilde{x}}_{j})$ is a kernel (function), $\bm{\tilde{x}}_i$ and $\bm{\tilde{x}}_j$ are respectively the data vectors for the $i$th and $j$th elements of $\bm x$ in the training dataset $\bm{\tilde{X}}$ of $\bm x$ obtained from other sources, eg MR anatomical images of the same (class of) subject/s, $\phi(\cdot)$ is a nonlinear function mapping $\bm x$ to a higher dimensional feature space but needs not to be explicitly known, and $\left \langle\cdot, \cdot \right\rangle$ denotes the inner product.

Typical kernels include the radial Gaussian kernel and the polynomial kernel
\begin{equation}
\label{Gaus}
k(\bm{\tilde{x}}_{i}, \bm{\tilde{x}}_{j})=\exp \left(-\|\bm{\tilde{x}}_{i}-\bm{\tilde{x}}_{j}\|^{2}/2 \sigma^{2}\right),
\end{equation}
\begin{equation}
\label{Poly}
k(\bm{\tilde{x}}_{i}, \bm{\tilde{x}}_{j}) = (\bm{\tilde{x}}_{i}^{T} \bm{\tilde{x}}_{j}+ \gamma)^{d},
\end{equation}
where $\sigma$, $\gamma$ and $d$ are kernel parameters, and $\|\cdot\|$ is the 2-norm of a vector.
For computation simplicity and performance improvement, the kernel matrix $\bm K$ in practice is often constructed using
\begin{equation}
\label{KCONS}
K(i,j)=\left\{\begin{array}{l}
k(\bm{\tilde{x}}_{i}, \bm{\tilde{x}}_{j}), \,\, j \in \cal J \\
0,  \text { otherwise }
\end{array}\right.
\end{equation}
where  $\cal{J}$ is the neighborhood of the $i$th element of emission image $\bm{x}$, with the neighborhood size $J$. Other methods such as $\bm{\tilde{x}}_j \in kNN$ (the $k$ nearest neighbor) of $\bm{\tilde{x}}_{i}$, measured by $\|\bm{\tilde{x}}_{i}-\bm{\tilde{x}}_{j}\|$, can also be used \cite{wang2015}.

Because the kernel matrix $\bm K$ constructed in (\ref{KCONS}) involves only a single kernel $k(\bm{\tilde{x}}_{i}, \bm{\tilde{x}}_{j})$, we will call it the single-kernel matrix hereafter.

Using the kernel space image model (\ref{KIM}), we can convert the forward projection (\ref{FPROJ}) and the log-posterior probability maximization (\ref{LMAP}) to their respective kernel space equivalents
\begin{equation}
\bar{\bm p}=\bm{H} \bm {K a} + \bm{r},
\end{equation}
\begin{equation}
\hat{\bm{a}}=\arg \max _{\bm{a} \geq 0} \,\, (L(\bm{p} |\bm{Ka})-\sum_{i = 1}^{m}\beta_{i}v_{i}(\bm{Ka})),
\end{equation}
and obtain a single-kernel regularized EM (KREM) algorithm
\begin{equation}
\label{KRMAP}
\begin{aligned}
\bm a^{n+1}=(\bm a^{n} \circ \bm{K}^{T} \bm{H}^{T}( \bm p \oslash (\bm H \bm K\bm a^{n}+\bm r)))\oslash \\
\bm{K}^{T}(\bm{H}^{T} \bm{1}_{M}+ \sum_{i = 1}^{m}\beta_{i}\frac{\partial V_{i}(\bm a^{n})}{\partial \bm a^{n}}),
\end{aligned}
\end{equation}
where $V_{i}(\bm a^{n}) = v_{i}(\bm{Ka}^{n})$ with $\bm K$ absorbed into its parameters. We can then use (\ref{KRMAP}) to estimate the coefficient $\bm a$ and use $\bm x^n = \bm {K a}^n$ to obtain an estimate of the image $\bm x$.

The KREM algorithm (\ref{KRMAP}) includes the KEM algorithm of \cite{wang2015, wang2019} as a special case when all $\beta_i = 0$. As shown in Sections \ref{Ex-results} and \ref{invivo}, the KEM algorithm tends to have larger reconstruction error and is sensitive to the number of iterations and the neighborhood size $J$ used in (\ref{KCONS}) to construct the single-kernel matrix $\bm K$. These problems may significantly degrade KEM's performance in PET image reconstruction, especially at the low counts. Whereas, the KREM algorithm (\ref{KRMAP}) with regularizations can effectively overcome these problems.

\subsection {PET Image Reconstruction with Multi-kernels and Multi-Regularizers }

The single-kernel matrix $\bm K$ in (\ref{KRMAP}) is the core of KREM algorithm. Its effect can be tuned by the selection of kernel $k(\bm{x}_i,\bm{x}_j)$, kernel parameters, and neighborhood size $J$ in (\ref{KCONS}). However, the efficacy of such tuning is limited as a single kernel is often insufficient for complex problems, such as heterogeneous data \cite{onen2011}. To solve this problem, a multi-kernel approach has been developed in machine learning, which combines multiple base kernels, such as that of (\ref{Gaus}) or (\ref{Poly}), to construct more complex kernel $k_M(\bm{x}_i,\bm{x}_j)$ and corresponding multi-kernel matrix $\bm K_M$ for significantly improved performance \cite{cortes2009, onen2011}.

Inspired by the success of multi-kernel approach in machine learning, we introduce this approach in PET image reconstruction. We construct the multi-kernel matrix $\bm K_M$ by
\begin{eqnarray}
\label{MK}
\bm{K}_M = [K_M(i,j)] = \prod_{l=1}^G \bm{K}_l = \prod_{l=1}^G [K_l(i,j)] \\
i,j = 1,2, \cdots, M, \,\, G \geq 1 \nonumber
\end{eqnarray}
and replace the single-kernel matrix $\bm K$ with $\bm K_M$ in KREM algorithm (\ref{KRMAP}) to derive the multi-kernel KREM (MKREM) algorithm for improved performance.

In (\ref{MK}), $\bm{K}_l = [K_l(i,j)] = [ k_l(\bm{x}_i,\bm{x}_j)]$, $i,j = 1,2, \cdots, M$, are constructed using (\ref{KCONS}) with different or same $k(\bm{x}_i,\bm{x}_j)$ having different or same parameters and neighborhood size $J$ for different  $l \in \{1, 2, \cdots, G\}$; and $\bm{K}_M = [K_m(i,j)]$, $i,j = 1,2, \cdots, M$, is the product of $\bm{K}_l$'s thus constructed.

The multi-kernel matrix $\bm{K}_M$ constructed using (\ref{MK}) includes single-kernel matrix $\bm{K}$ as a special case when $ G =1$. Hence, the MKREM algorithm derived from replacing $\bm{K}$ by $\bm{K}_M$ in (\ref{KRMAP}) includes KREM algorithm as a special case. We therefore call (\ref{KRMAP}) the MKREM algorithm and KREM algorithm alternately when there is no confusion.

It can be readily proven that for all $i,j \in \{1,2, \cdots, M\}$ the $ij$th element of $\bm{K}_M$, $K_M(i,j) = k_M(\bm{\tilde{x}}_i, \bm{\tilde{x}}_j)$, is a new kernel resulting from multiplications and summations of the base kernels $k_l(\bm{\tilde{x}}_i, \bm{\tilde{x}}_j)$, $l =1, 2, \cdots,G$. Therefore, $\bm{K}_M$ constructed from (\ref{MK}) is indeed a multi-kernel matrix \cite{cortes2009, onen2011}. The proof is omitted due to space limit. 

\subsection{Kernel Space Image Dictionary Regularizer}
\label{RKD}
The regularizers $V_i(\bm a^n)$ in (\ref{KRMAP}) are symbolic. In this and the next subsections, we will present two specific regularizers to be used in this work. To facilitate presentation, we introduce the kernel space image dictionary for PET emission image $\bm x$.

Let $\bm {K}_{al}$, $\bm {K}_{bl} \in R^{M \times M}$, $l = 1, 2, \cdots, G$, be two groups of single-kernel matrices constructed using (\ref{KCONS}) with the kernel $k_l(\cdot,\cdot)$ and neighborhood size $J = J_{al}$ for $\bm {K}_{al}$ and $J = J_{bl}$ for $\bm {K}_{bl}$. Let $\bm {K}_{Ma}$, $\bm {K}_{Mb} \in R^{M \times M}$ be two multi-kernel matrices constructed using (\ref{MK}) with $\bm {K}_{al}$ for $\bm {K}_{Ma}$ and $\bm {K}_{bl}$ for $\bm {K}_{Mb}$. Here $\bm {K}_{Ma}$ is used as $\bm {K}_{M}$ in MKREM algorithm (\ref{KRMAP}) and $\bm {K}_{Mb}$ is used in the dictionary learning described below.
When $G = 1$, 
$\bm {K}_{Ma} = \bm {K}_{a1} := \bm {K}_{a}$ and $\bm {K}_{Mb} = \bm {K}_{b1} := \bm {K}_{b}$ reduce to the single-kernel matrices.
As shown by the experiment results in Section \ref{Ex-results}, $J_b < J_a$ and hence $\bm {K}_{Mb} \neq \bm {K}_{Ma}$ are generally needed for the best reconstruction performance.

Since $\bm {K}_{Ma}$, $\bm {K}_{Mb} \in R^{M \times M}$ and $\bm {K}_{Mb} \neq \bm {K}_{Ma}$, $\bm {K}_{Mb}$ admits the kernel matrix factorization
\begin{equation}
\bm K_{Mb} = \bm K_{Ma} \bm{\tilde{K}},  \,\, \bm{\tilde{ K}} \in R^{M \times M}.
\end{equation}
Replacing $\bm {K}$ in (\ref{KIM}) by $\bm {K}_{Ma}$ and $\bm K_{Mb}$ respectively gives the representations of emission image $\bm x$ under the multi-kernel matrices $\bm {K}_{Ma}$ and $\bm K_{Mb}$
\[
\bm x = \bm K_{Ma}\bm a = \bm K_{Mb} \bm b = \bm K_{Ma} \bm{\tilde{ K}} \bm b,
\]
where $\bm a, \bm b \in R^M$ are respectively the coefficient vectors of $\bm x$ under $\bm {K}_{Ma}$ and $\bm K_{Mb}$ and are related by $\bm a =\bm{\tilde{ K}} \bm b$.

The $\bm a$ and $\bm b$ can be viewed as kernel space image signals and hence both admit dictionary representation \cite{elad2010}. Defining a dictionary $\bm D_b \in R^{M \times S}$ with $S$ $(S \geq M)$ dictionary atoms and a coefficient vector $\bm c \in R^S$ for $\bm b$, we can represent $\bm b$ and $\bm a$ respectively as
\[
\bm{b} = \bm{D}_b \bm c, \,\,
\bm{a} = \bm{\tilde{ K}} \bm b =  \bm{D}_a \bm c,
\]
where $\bm{D}_a := \bm{\tilde{ K}}\bm{D}_b$ is a dictionary for  $\bm a$. It can be seen that  $\bm{\tilde{ K}}$ is a mapping of $\bm b$ and $\bm{D}_b$ in the kernel space of $\bm K_{Mb}$ to $\bm a$ and $\bm{D}_a$ in the kernel space of $\bm K_{Ma}$.

Let $\bm{\tilde{X}} = [\bm{\tilde{x}}_1, \cdots, \bm{\tilde{x}}_i, \cdots, \bm{\tilde{x}}_T] \in R^{M \times T}$ be the matrix consisting of $T$ training data vectors $\bm{\tilde{x}}_i$ for $\bm x$, obtained from other sources, eg MR anatomical images of the same (class of) subject/s. We use the following sparse dictionary learning to obtain the dictionary $\bm D_b$ \cite{elad2010}
\begin{equation}
\label{DL}
\underset{\bm D_b, \bm C}{\operatorname{argmin}} \,\, \| \bm B -\bm D_b \bm C\|_F^{2}, \text { s.t. }\left\|\bm c_i\right\|_0 \leq s,\end{equation}
where $\bm B := \bm K_{Mb}^{-1} \bm{\tilde X}$ is the data matrix for dictionary learning, $\bm C := [\bm c_1, \cdots, \bm c_i, \cdots, \bm c_T]$ is the matrix consisting of $T$ estimates of the coefficient vector $\bm c$, $\|\cdot\|_F$ is the Frobenius norm, and $s$ is a sparsity constraint constant. Theoretically, $\bm K_{Mb}$ is always positive semidefinite \cite{taylor2004}. In the rare case it is singular, $(\bm K_{Mb}+ \epsilon I)^{-1}$, with  $\epsilon > 0$ a small constant, can be used as $\bm K_{Mb}^{-1}$ without affecting image reconstruction result.

In case the number of training data vectors $T$ is limited, eg $T = 1$, or the row dimension of $\bm B$ is too high, we can use the patch-based dictionary learning to increase training data or reduce computation complexity \cite{elad2010}
\begin{equation}
\label{PDL}
\underset{\bm D_b, \bm C}{\operatorname{argmin}} \,\, \|\mathbb{P}(\bm B)-\mathbb{P}(\bm D_b \bm C)\|_F^{2}, \text { s.t. }\left\|\bm c_{i}\right\|_{0} \leq s,
\end{equation}
where $\mathbb{P}$ is the operator extracting patches from $\bm B$ and $\bm D_b \bm C$.

A number of well developed methods can be used to solve (\ref{DL}) ((\ref{PDL})) for $\bm D_b$ and $\bm C$ \cite{elad2010}. All these methods are based on the alternating optimization: 1) fix $\bm D_b$ and solve (\ref{DL}) ((\ref{PDL})) for $\bm C$, 2) fix $\bm C$ and solve (\ref{DL}) ((\ref{PDL})) for $\bm D_b$, 3) repeat 1) and 2) until the maximum number of iterations is reached. In this work, we use the orthogonal matching pursuit (OMP) in 1) to compute $\bm C$ and the K-SVD in 2) to compute $\bm D_b$ \cite{elad2010}.

With above learned dictionary $\bm D_b$, we can obtain
\begin{equation}
\bm{D}_a = \bm{\tilde{ K}}\bm{D}_b
\end{equation}
and introduce the first regularizer and its derivative
\begin{equation}
\label{V1}
V_{1}(\bm{a}^n) := \frac{1}{2}\|\bm{a}^{n} -  \bm D_a \bm{c}^{n}\|^{2},
\end{equation}
\begin{equation}
\label{DV1}
\frac{\partial V_{1}(\bm{a}^n)}{\partial \bm{a}^{n}} = (\bm{a}^{n} - \bm D_a \bm{c}^{n}),
\end{equation}
where $\bm{a}^n$ is from the $n$th MKREM iteration in (\ref{KRMAP}), and $\bm c^n$ is from solving the optimization below for $\bm c^n$ using $\bm{a}^n$ and OMP before the $(n+1)$th MKREM iteration.
\begin{equation}
\label{cn}
\min_{\bm c^n}\|\bm{a}^{n} -  \bm D_a \bm{c}^{n}\|^{2}.
\end{equation}

\subsection{Kernel Space Graph Laplacian Quadratic Regularizer}
\label{GQ}

As mentioned above, the neighborhood size $J_b < J_a$ is generally needed for the multi-kernel matrix $\bm K_{Mb}$ used in dictionary learning. This is to allow the dictionary to capture the local texture details embedded in the training image data which is usually the images with anatomical details. To avoid over-localization of the image reconstruction, we treat the image data as the signal on graph and use the graph Laplacian quadratic \cite{shuman 2013} of the estimated kernel image $\bm a^n$, given in (\ref{V2}), as the second regularizer to smooth the estimated PET image $\bm x^n = \bm{K}_{Ma} \bm{a}^n$ on the image graph 
at each iteration of MKREM algorithm (\ref{KRMAP}).
\begin{equation}
\label{V2}
V_{2}(\bm a^n) =  \frac{1}{2}\bm{a}^{nT} \bm{K}_{Ma}^T \bm Q \bm{K}_{Ma} \bm{a}^n
\end{equation}
with the derivative
\begin{equation}
\label{DV2}
\frac{\partial V_{2}(\bm{a}^n)}{\partial \bm{a}^n} = \bm{K}_{Ma}^T\bm{Q}\bm{K}_{Ma}\bm{a}^n =  \bm{Q}_a\bm{a}^n,
\end{equation}
where $\bm{Q}_a := \bm{K}_{Ma}^T\bm{Q}\bm{K}_{Ma}$ and $\bm Q$ is the Laplacian matrix of the graph of training image data  described below.

Inspired by \cite{dijk2017}, we treat a training image $\bm{\tilde{x}} = [\tilde{x}_1, \cdots, \tilde{x}_i,\cdots, \tilde{x}_M]^T$ as a graph signal and construct its Laplacian matrix as follows:
We first use a window of size $m$ centered at the $i$th element $\tilde{x}_i$ of $\bm{\tilde{x}}$ to extract $m$ elements and use them to form the vector $\bm{y}_i$, for $i = 1, 2, \cdots, M$. Then we construct a distance matrix $\bm E$ with elements $E(i,j) = \|\bm{y}_i -\bm{y}_j\|$, $i, j = 1, 2, \cdots, M$, and convert $\bm E$  to an affinity matrix $\bm W$ by an adaptive Gaussian function $W(i,j) = exp(-E(i,j)/b_i)^2$, where $b_i$ is the ($knn+1$) smallest $E(i,j)$ in the $i$th row of $\bm E$. Next, we symmetrize $\bm W$ to $\bm{\tilde{W}} = \bm W + \bm{W}^T$ and construct a Markov transition matrix $\bm Z$ with $Z(i,j) = \tilde{W}(i,j)/\sum_{j}\tilde{W}(i,j)$. If there are $T > 1$ training images available, we apply above procedure to each image $\bm{\tilde{x}}_i$ to get $\bm Z_i$ and take $\bm Z =\frac{1}{T} \sum_{i = 1}^T \bm Z_i$.

Using above constructed $\bm Z$, we construct the Laplacian matrix
$\bm {Q} = \bm{I} - \bm{Z}^t,$
where $\bm I$ is identity matrix and $\bm Z^t$ is the optimal Markov transition matrix.  The optimal power $t$ of $\bm Z^t$ is determined by
$\frac{||\bm{Y} \bm{Z}^t - \bm{Y} \bm{Z}^{t-1}||_F^2} {||\bm{Y} \bm{Z}^{t-1}||_F^2} \leq \epsilon,$
with $\epsilon > 0$ a small constant, $\bm{Y} := [\bm y_1, \cdots, \bm y_i, \cdots, \bm y_M]^T$ and $\bm y_i$ as described above.

\subsection{Proposed Method}

Substituting $\partial V_{1}(\bm{a}^n)/\partial \bm{a}^n$ and $\partial V_{2}(\bm{a}^n)/\partial \bm{a}^n$  derived in (\ref{DV1}) and (\ref{DV2}) into (\ref{KRMAP}), we obtain the full mathematical expression of MKREM algorithm proposed for PET image reconstruction
\begin{equation}
\label{Imgrecon}
\begin{aligned}
\bm a^{n+1}=\{\bm a^{n} \circ \bm{K}_{Ma}^{T} \bm{H}^{T}[ \bm P \oslash (\bm H \bm K_{Ma} \bm a^{n}+\bm r)]\}\oslash\\ \bm{K}_{Ma}^{T}[\bm{H}^{T} \bm 1_{M}+\beta_1(\bm{a}^{n} - \bm D_a \bm{c}^{n}) + \beta_2\bm {Q}_{a}\bm{a}^n],
\end{aligned}
\end{equation}
where $\bm{K}_{Ma}$ has replaced $\bm K_M$ in (\ref{KRMAP}) as explained in Subsection \ref{RKD}. When $\bm{K}_{Ma}$ is constructed with $G=1$ in (\ref{MK}),  it reduces to a single-kernel matrix $\bm K_a$ and (\ref{Imgrecon}) becomes the full mathematical expression of KREM algorithm (\ref{KRMAP}). The pseudo-code of the proposed method is given in Algorithm 1.

\begin{algorithm}[htb]
	\caption{Pseudo-code of the proposed method}
	\label{alg:Framwork3}
	\begin{algorithmic}[1]
		\Require
		PET projection data vector $\bm p$; 
		PET system matrix $\bm{H}$;
		training image data $\bm{\tilde{X}}$; 
		initial value $\bm{a}^0$ for kernel image $\bm{a}$; maximum iteration number $n_{max}$ for MKREM algorithm (\ref{Imgrecon}).
		\Ensure
		Reconstructed  PET image $\bm{x}$ 
		\State Extract $\bm{\tilde{x}_i}$'s from $\bm{\tilde{X}}$; specify $G$, $J_{al}$, $J_{bl}$ and $k_l(\cdot,\cdot)$ for $l = 1, 2,\cdots, G$; construct single-kernel matrices $\bm{K}_{al}$ ($\bm{K}_{bl}$) using $\bm{\tilde{x}_i}$'s, $k_l(\cdot,\cdot)$, $J_{al}$ ($J_{bl}$) and (\ref{KCONS}); construct multi-kernel matrix $\bm{K}_{Ma}$ ($\bm{K}_{Mb}$) using $\bm{K}_{al}$ ($\bm{K}_{bl}$), $G$ and (\ref{MK}).			
		\State Generate dictionary learning data $\bm B = \bm{K}_{Mb}^{-1}\bm \tilde X$; use $\bm B$ in (\ref{DL}) ((\ref{PDL})) to learn dictionary $\bm D_b$ by OMP and K-SVD; factorize $\bm{K}_{Mb} = \bm{K}_{Ma}\bm{\tilde{K}}$ to get dictionary $\bm D_a = \bm{\tilde{K}}\bm D_b$.		
		\State Use $\bm{\tilde{x}_i}$'s from Step 1 and procedure in Subsection \ref{GQ} to construct Laplacian matrix $\bm {Q}$ and $\bm {Q}_a = \bm{K}_{Ma}^T\bm {Q}\bm{K}_{Ma}$.
		\State $n = 0$
		\While {$n < n_{max}$}.						
		\State Solve (\ref{cn}) for $\bm{c}^n$ using  $\bm a^n$, $\bm D_a$ and OMP.
		\State Compute $\bm {a}^{n+1}$ using (\ref{Imgrecon}) with $\bm a^n$, $\bm{c}^n$, $\bm D_a$ and $\bm Q_a$.
		\State $n = n + 1$.
		\EndWhile
		\State Compute $\bm x = \bm K_{Ma} \bm a^{n_{max}}$ to get PET image $\bm{x}$.		
	\end{algorithmic}
\end{algorithm}

\section{Simulation studies} 
\label{Ex-results}

To investigate their behavior and performance, we have applied the proposed MKREM and KREM algorithms to the simulated low count and high count sinogram data to reconstruct the PET images and compare them with the existing PET image reconstruction algorithms. Since the KEM algorithm \cite{wang2015} has proven to be superior to many state-of-the-art PET image reconstruction algorithms, we have mainly compared the MKREM, KREM and KEM algorithms. We have also compared with MLEM with Gaussian filter (MLEM+F) to show some common problems of PET image reconstruction methods that are based on the un-regularized EM. 

\subsection{PET and Anatomical MRI Data Simulation}

The same method as that of \cite{wang2015} was used to generate 3D PET activity images. The 3D anatomical model from BrainWeb database consisting of 10 tissue classes was used to simulate the corresponding PET images \cite{cocosco1997},\cite{collins1998}. The regional time activities at 50 minutes were assigned to different brain regions to obtain the reference PET image. The matching T1-weighted 3D MR image generated from BrainWeb database was used as training image data. A slice of the anatomical model and the reference PET image are shown in Fig. 2 (a) and (b), respectively. The bright spot in the periventricular white matter is a simulated lesion created by BrainWeb.

The reference PET image was forward projected to generate noiseless sinograms using the PET system matrix generated by Fessler toolbox. The count number of noiseless sinograms was then set to 370k and 64k, respectively, to obtain the noiseless high count sinograms and the noiseless low count sinograms. The thinning Poisson process was applied to the noiseless high count and low count sinograms to generate the noisy high count and noisy low count sinograms for image reconstruction tests. Attenuation maps, mean scatters and random events were included in all reconstruction methods to obtain quantitative images. Ten noisy realizations were simulated and each was reconstructed independently for comparison.

To simulate the real scanning process and test the robustness of different methods, the BrainWeb generated MR images with noise were used as the training data $\bm \tilde{X}$ to construct the kernel matrices, dictionaries and Laplacian matrix of the proposed MKREM and KREM methods.

\begin{figure}[!t]
	\centerline{\includegraphics[width=0.52 \columnwidth,height=0.41 \linewidth]{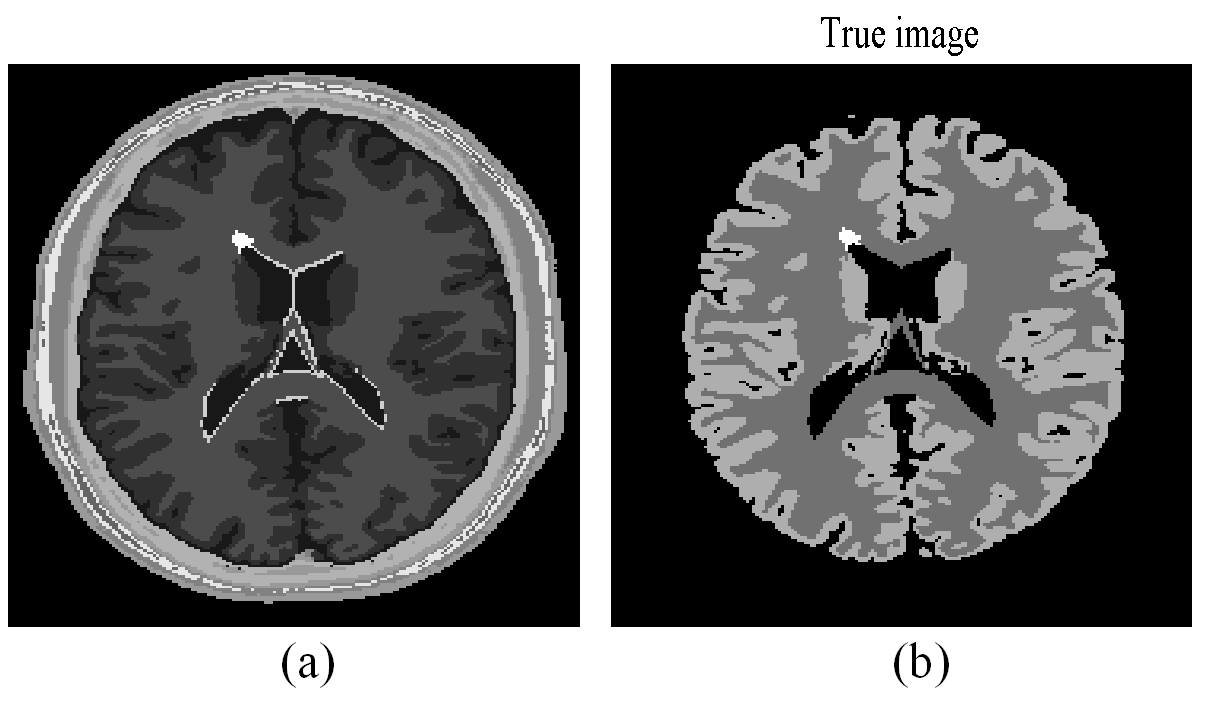}}
	\caption{ (a) A slice of anatomical model.
		(b) The corresponding simulated reference PET image. The bright spot is a lesion simulated and inserted by BrainWeb.}
	\label{fig3}
\end{figure}

\subsection{Algorithms Setup and Performance Metrics}

We used $G = 2$, $\bm K_l \equiv \bm K_{a}, \forall l$, for $\bm K_{Ma}$ and $\bm K_l \equiv \bm K_{b}, \forall l$, for $\bm K_{Mb}$ in (\ref{MK}) to construct the multi-kernel matrices $\bm K_{Ma} = \bm K_{a}\bm K_{a}$ for the MKREM algorithm (\ref{KRMAP}) and $\bm K_{Mb} = \bm K_{b}\bm K_{b}$ for dictionary learning. The above single-kernel matrices $\bm K_{a}$ and $\bm K_{b}$ were constructed using the Gaussian kernel (\ref{Gaus}) with $\sigma = 0.5$, $J_a = 21$ and $J_b =3$ in (\ref{KCONS}). The $\bm K_{a}$ thus constructed was used as the kernel matrix $\bm K$ in the KREM algorithm (\ref{KRMAP}).

For kernel space dictionary learning, the training data $\bm{B} = K_{Mb} \bm \tilde{X}$ was divided into 12960 overlapping 5 $\times$ 5 patches. To ensure the generalization and efficiency of learned dictionary,  400 patches were randomly chosen from the 12960 patches. The matrix of patched training data $\mathbb{P}(\bm{B})$ was $25\times400$. The  number of dictionary atoms was set to 50 to obtain over-complete dictionaries $\bm D_b$ and $\bm D_a$. The sparsity constraint constant $s =50$ and the total iteration number = 50 were used in dictionary learning.

The data window of size $m = 9$ was used to construct the Markov transition matrix $\bm Z$, and $\epsilon = 10^{-4}$ was used to find the optimal $t = 3$ for the Laplacian matrix $\bm Q = \bm I - \bm Z^t$.

For the low count image reconstruction, we tuned the weighting parameters to $\beta_1 = 0.003$ and $\beta_2 = 0.13$, and for the high count image,  $\beta_1= 0.001$ and $\beta_2 = 0.04$.

The same $\bm K_a$ as described above was used as the kernel matrix for the KEM algorithm. The post-filter of MLEM was Gaussian filter with window size 5$\times$5.

For fair comparison, the total iteration numbers and parameters of the compared methods were tuned separately to achieve their respective minimum mean square error (MMSE) in image reconstruction.


The mean square error (MSE), $\text{MSE}(\bm x, \bm f)=\frac{\sum_{j}^{M}\left(\bm x_{j}-\bm f_{j}\right)^{2}}{\sum_{j}^{M}(\bm f_{j})^2}$, the ensemble mean square bias, $\text {Bias}=\frac{\sum_{j}^{M}\left(\bm x_{j}-\bm f_{j}\right)}{\sum_{j}^{M}\left(\bm f_{j}\right)}$, and the standard deviation, $\operatorname{Var}={\frac{1}{N} \frac{\left.\sum_{i}^{O} \sum_{j}^{M}\left(\bm x_{j}^{i}-\frac{1}{O} \sum_{i}^{O} \bm x_{j}^{i}\right)\right)^{2}}{\sum_{j}^{M} \bm f_{j}^{2}}}$, were used for the quantitative performance evaluation and comparison of the algorithms,
where $\bm x_j$ is the $j$th element of reconstructed PET image, $\bm f_j$ is the $j$th element of the reference PET image (the true image) and $O$ is the total number of noisy realizations.

\subsection{Image Reconstruction Performance}

\begin{figure*}[ht] \centerline{\includegraphics[width=15cm,height=8cm]{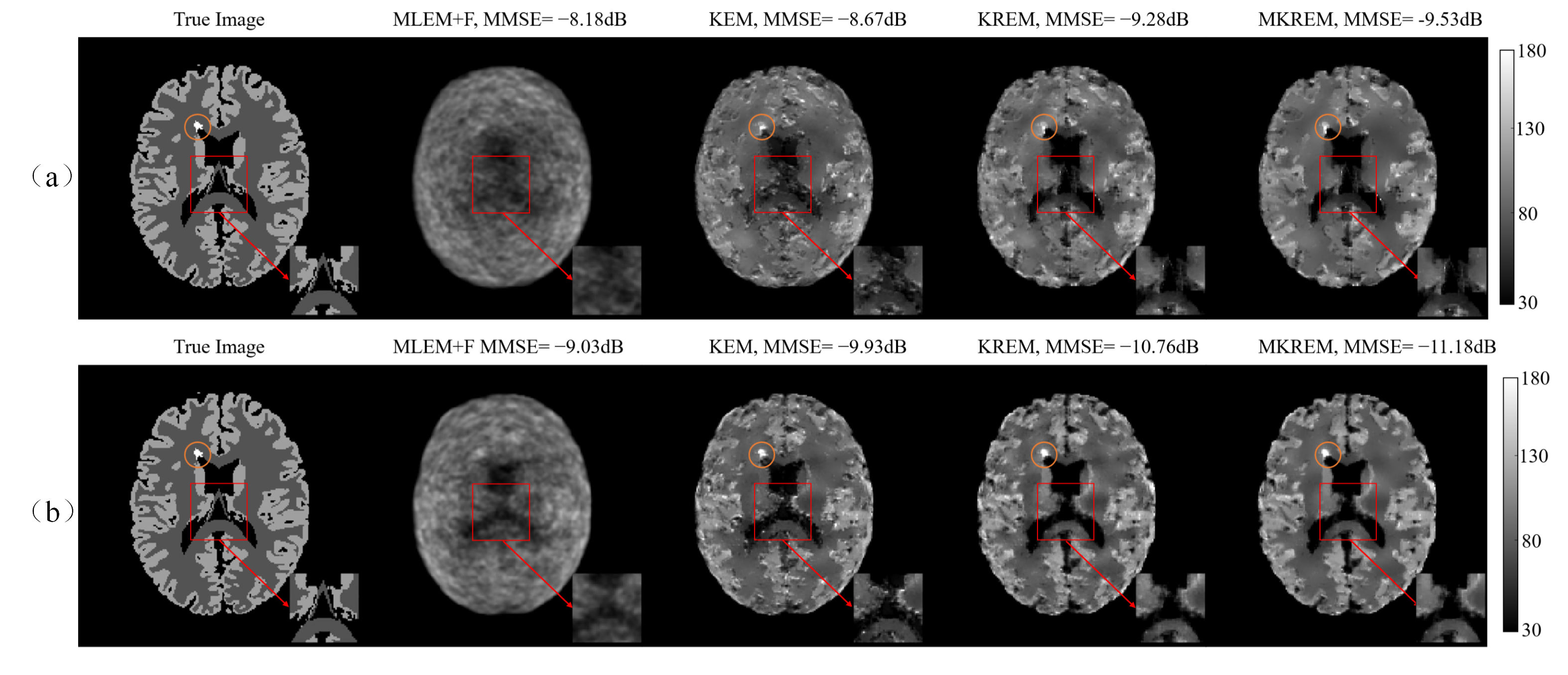}}
	\caption{Reconstruction results from simulated data. (a) Low count (64k) image. (b) High count (370k) image.}
	\label{fig4}
\end{figure*}

Fig. \ref{fig4}(a) shows the true image and the images reconstructed by different algorithms for the slice of low count image. 
As seen from the figure, KREM and MKREM achieve lower MMSE and better visual effects than those of MLEM+F and KEM, and the KREM and MKREM images appear to be less noisy, higher contrast and clearer texture. The intensity (brightness) of the lesion area in KREM and MKREM images, especially that of MKREM images, is closer to the true image than in KEM and MLEM images. As shown in the enlarged cutouts, the edges and texture of the true image are better preserved by KREM and MKREM than by MLEM+F and KEM.

\begin{figure}[t]
	\centerline{\includegraphics[width= 9cm,height = 10cm]{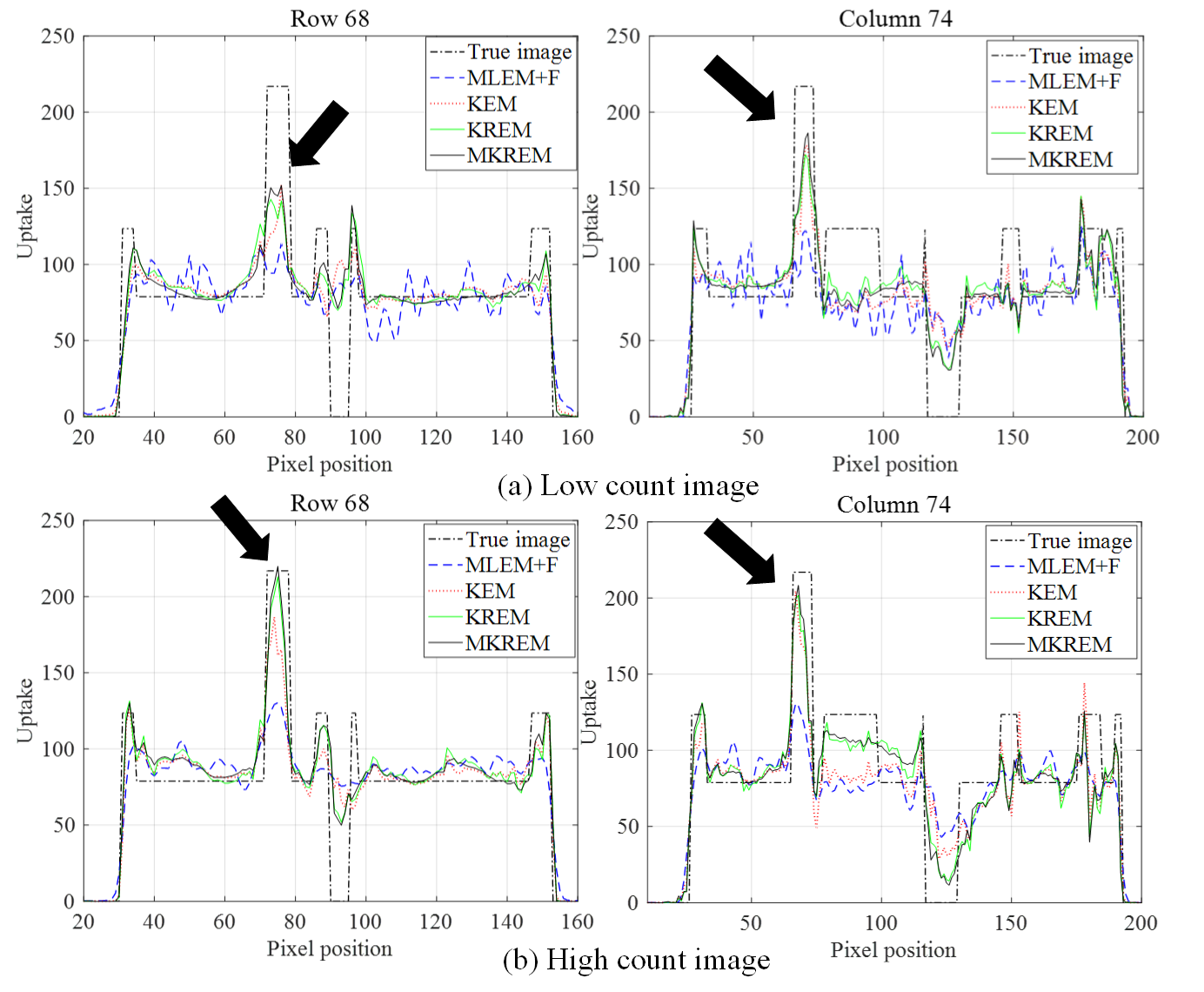}}
	\caption{Pixel line profile plots across the lesion in (a) low count image and (b) high count image.}
	\label{fig5}
\end{figure}

Fig. \ref{fig4}(b) shows the true image and the images reconstructed by different algorithms for the same slice of high count image. 
Due to the high count of this image, all algorithms achieve better results. But KREM and MKREM images are sharper and less noisy with lower MMSE and globally clearer image details than MLEM+F and KEM images. The lesion area in KREM and MKREM images is more pronounced
than in MLEM+F and KEM images. The quantitative comparison is given in Fig. \ref{fig5}. As shown in the enlarged cutouts, KREM and MKREM can more effectively preserve the boundaries and details of the image than MLEM+F and KEM.

Fig. \ref{fig5} plots the pixel line profiles across the lesion in the low count and high count images. The lesion area shows up as one peak indicated by the black arrow, which is significantly different from the white matter region in the high count image. Compared with other three methods, MKREM has the highest uptake in the lesion area in both low count and high count images. 

The above results show that KREM and MKREM outperforms MLEM+F and KEM for both low count and high count images in terms of image visual effect and reconstruction MMSE. 

Comparing the results in Fig.'s \ref{fig4}-\ref{fig5}, it can be clearly seen that MKREM further outperforms KREM in image reconstruction MMSE, image visual effects and pixel uptakes for the low count and high count images. 
\begin{figure}[t]
	\centerline{\includegraphics[width=7cm,height = 10cm] {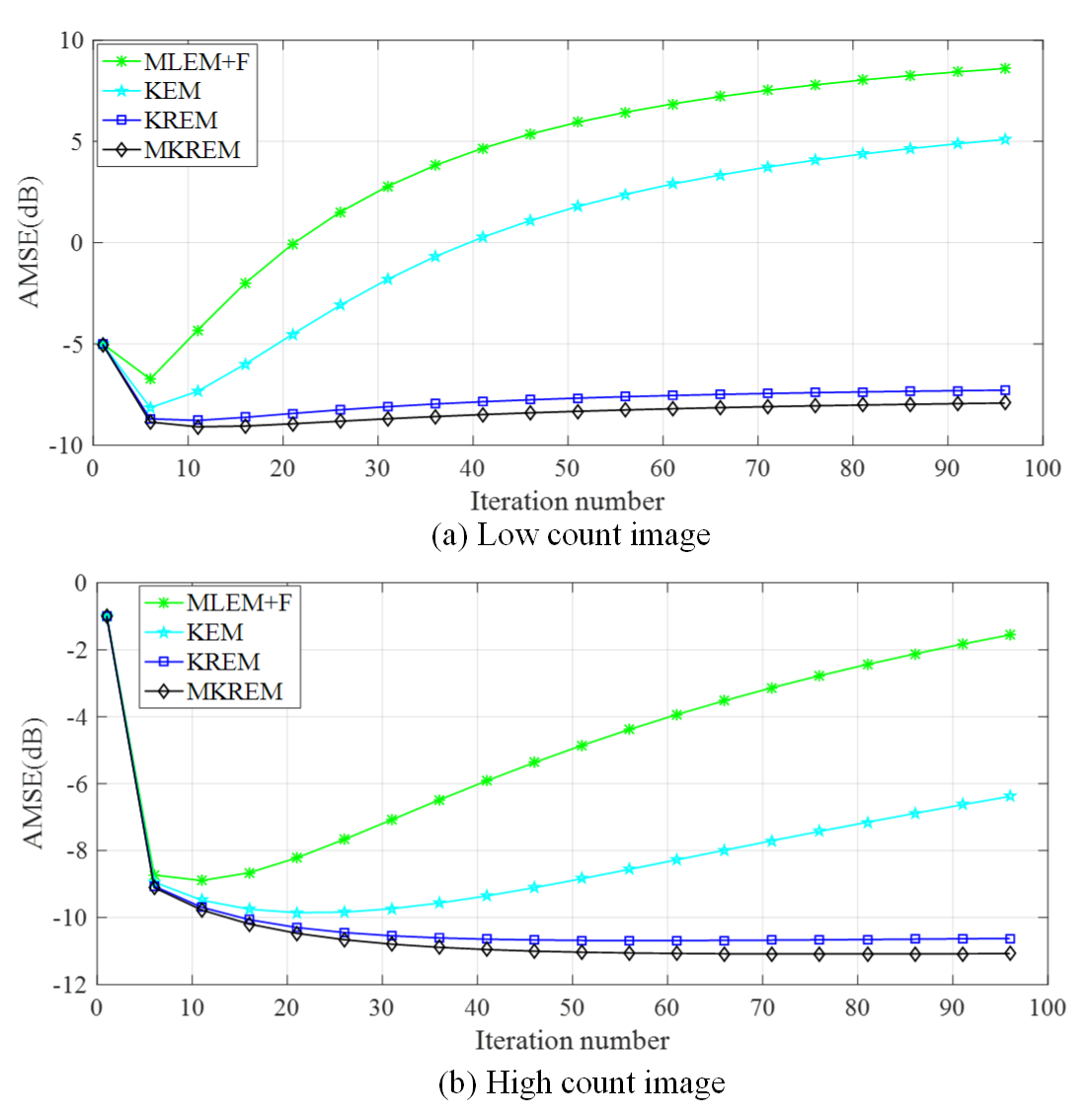}} 
	\caption{Average MSE versus iteration number of different algorithms for (a) low count image and (b) high count image.}
	\label{fig6}
\end{figure}

\subsection{Average MSE, Convergence Speed and Robustness}

Fig. \ref{fig6} plots the average MSE (AMSE) versus the iteration number of MLEM+F, KEM, KREM and MKREM over ten noisy realizations. As seen from the plotted curves, all these algorithms approach their respective minimum AMSE's rather quickly as the iteration number increases, but the minimum AMSE's of KREM and MKREM are lower than those of MLEM+F and KEM, and that of MKREM is the lowest. When the iteration number further increases to beyond the minimum AMSE points, the AMSE's of MLEM+F and KEM start to increase rapidly. As shown before in Fig. \ref{fig1}, such increase can last thousands iterations before the AMSE's slowly converge to some very high values, resulting in extremely noisy and useless images. The same phenomenon has been observed and analyzed in \cite{barrett1999} for non-kernelized MLEM in PET image reconstruction, and it is shown now for KEM in the same application. In stark contrast, the AMSE's of KREM and MKREM are quite stable, with negligible increases from their minimum AMSE's over thousands iterations. As shown in Fig. \ref{fig6}, these results hold for both low count and high count cases.

\begin{figure}[t]
	\centerline{\includegraphics[width=7cm,height = 10.5cm]{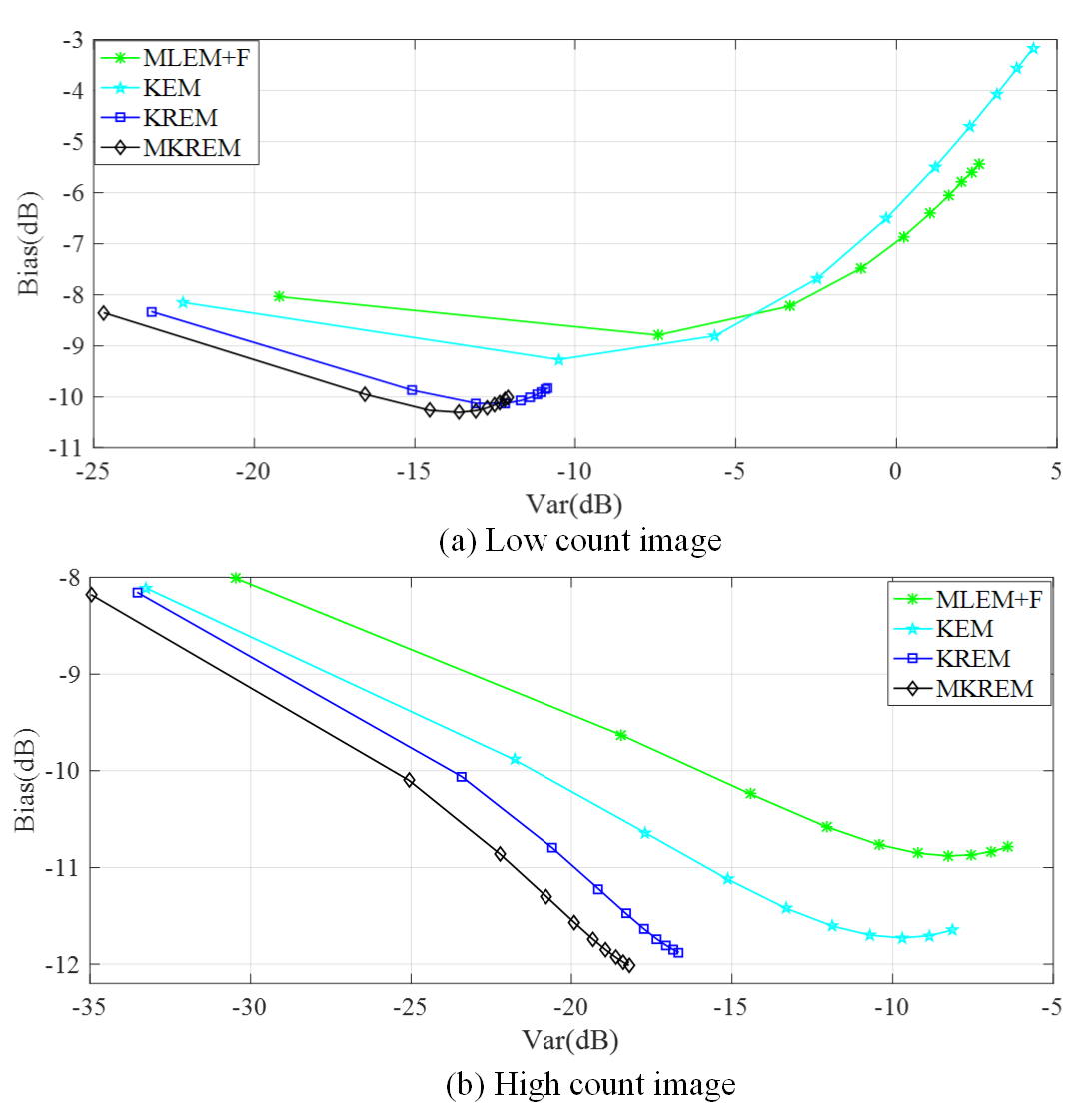}}
	\caption{Ensemble mean squared bias versus mean variance trade-off of different algorithms at the iterations 3 to 93 in increments of 10. (a) Low count image and (b) high count image.}
	\label{fig7}
\end{figure}


Fig. \ref{fig7} shows the ensemble mean squared bias versus mean variance trade-off of different algorithms at the iterations 3 to 93 in increments of 10.  As seen from the figure, the proposed KREM and MKREM always get much lower bias and variance simultaneously than those of MLEM+F and KEM for both low count and high count images.

\begin{figure}[ht]
	\centerline{\includegraphics[width=7cm,height = 10cm]{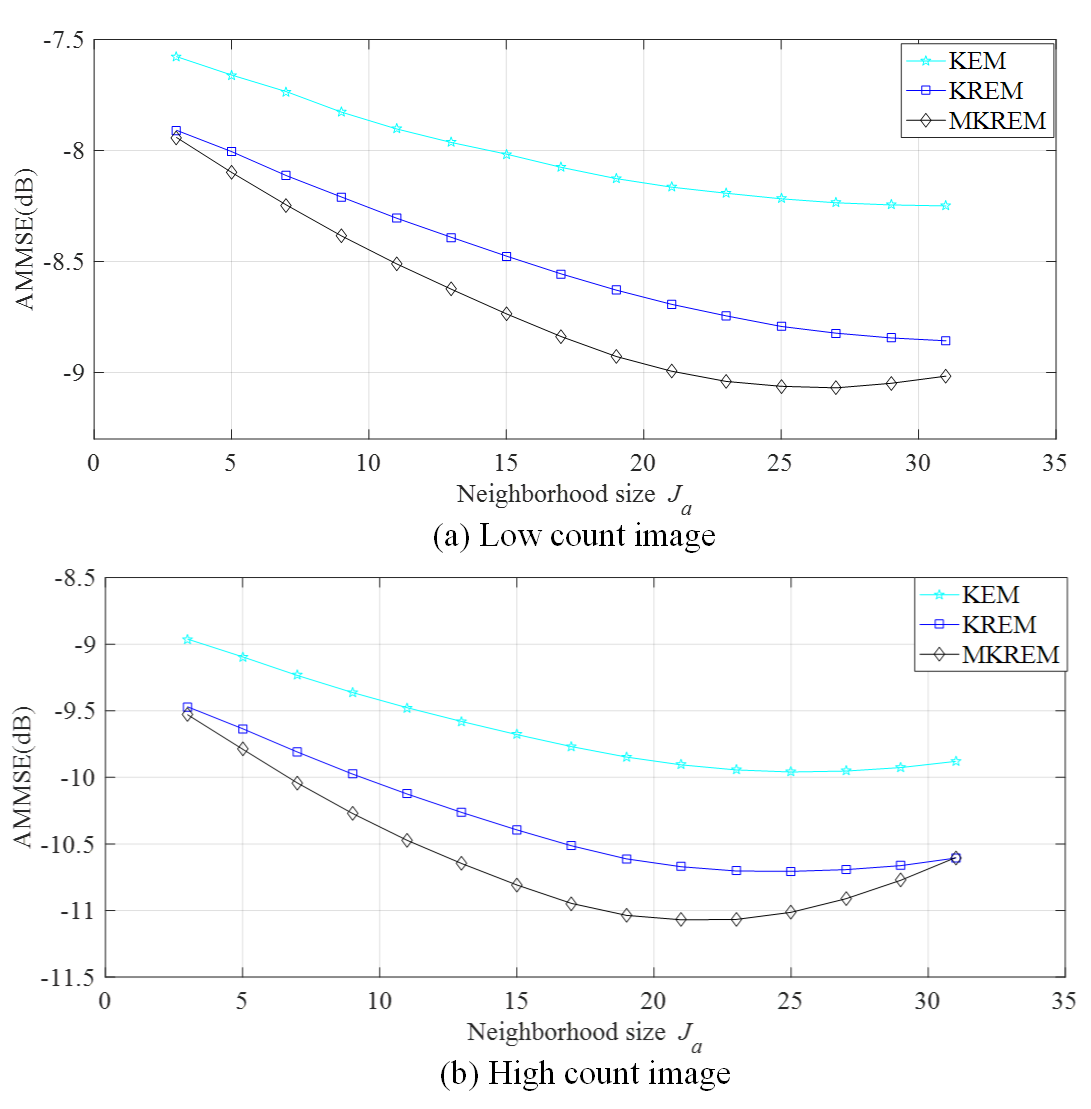}}
	\caption{AMMSE versus neighborhood size $J_a$ for different algorithms in (a) low count image and (b) high count image.}
	\label{fig8}
\end{figure}


Fig. \ref{fig8} shows the AMMSE's of KEM, KREM and MKREM versus the neighborhood size $J_a$ over ten noisy realizations. As seen from the figure, in both the low count and high count cases, KREM and MKREM achieve much lower AMMSE than that of KEM over a wide range of $J_a$ selections, and MKREM achieves the lowest. Hence, KREM and MKREM are less sensitive to the selection of $J_a$.

\begin{figure*}[!ht]
	\centerline{\includegraphics[width=18.5cm,height = 7cm]{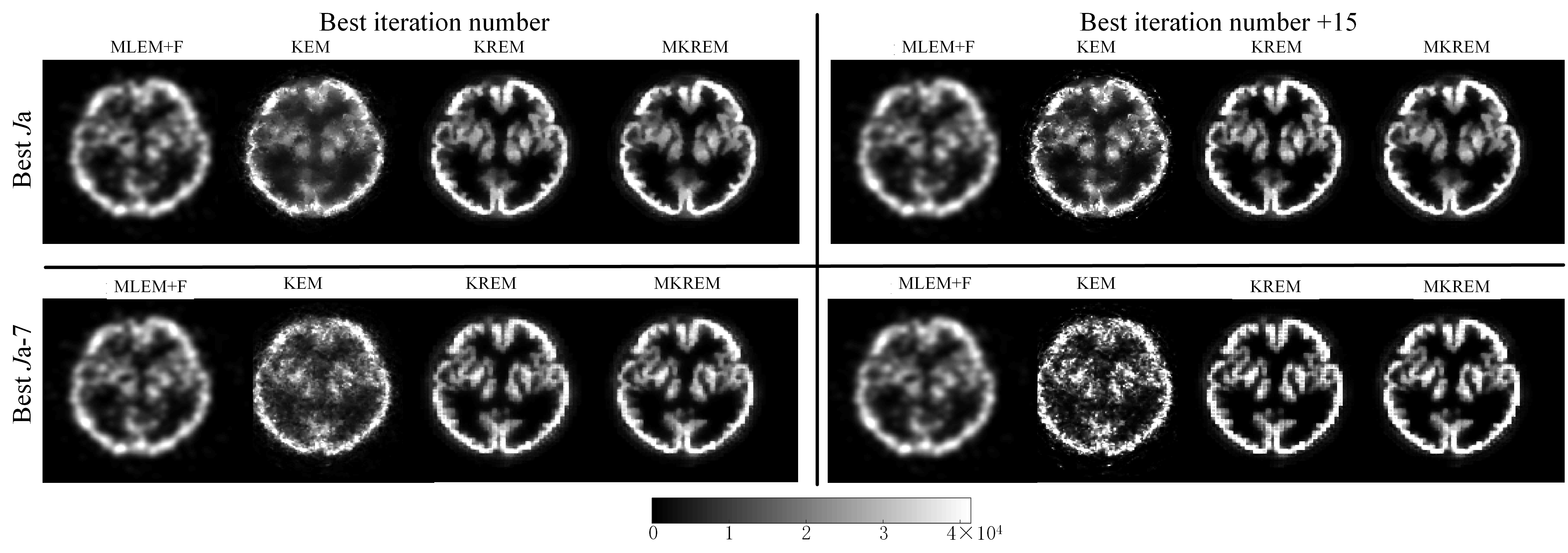}}
	\caption{Phantom data reconstruction result of all methods under varying neighborhood sizes and iteration numbers. }
	\label{fig10}
\end{figure*}

\begin{figure*}[!ht]
	\centerline{\includegraphics[width=18.5cm,height = 7cm]  {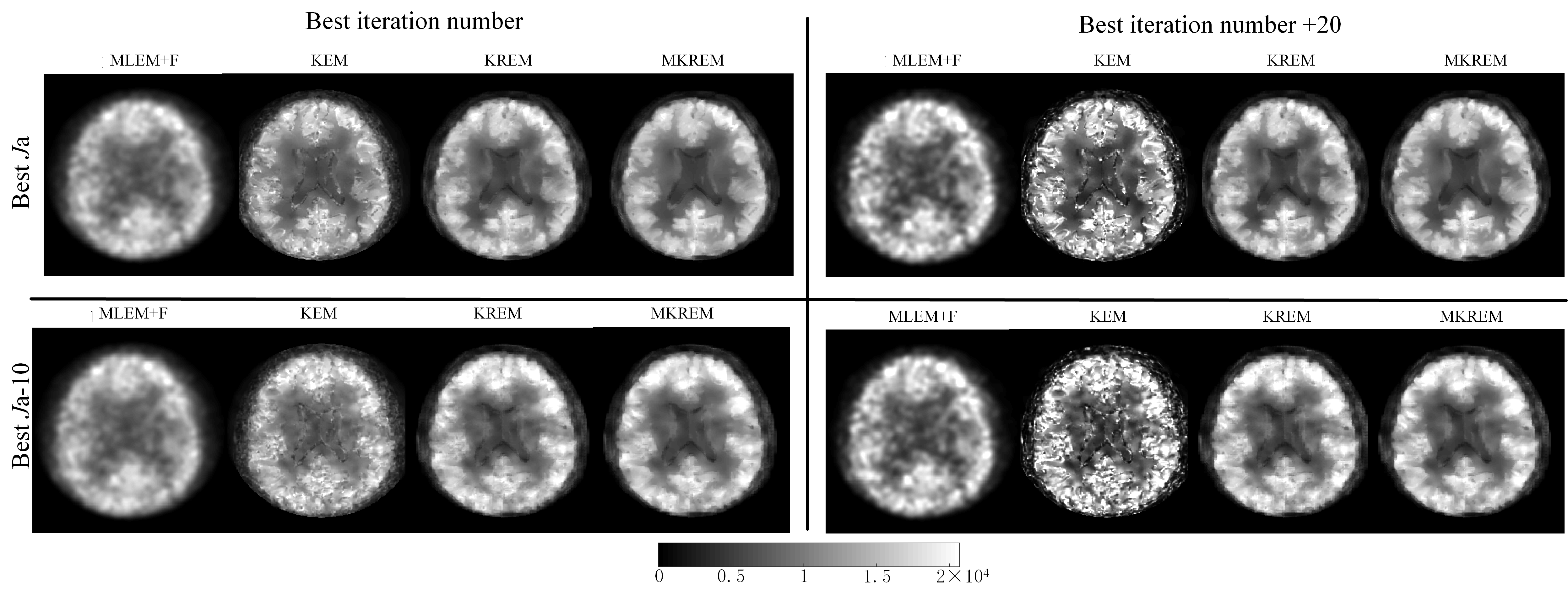}}
	\caption{In vivo human data reconstruction result of all methods under varying neighborhood sizes and iteration numbers}
	\label{fig11}
\end{figure*}
\section{Tests on In-vivo Data}
\label{invivo}
To validate the results of simulation studies and further compare the performance of MLEM+F, KEM, KREM and MKREM methods, we have applied them to in-vivo phantom and human data. The PET in-vivo phantom and human data and corresponding MR images were acquired from a Siemens (Erlangen) Biograph 3 Tesla molecular MR (mMR) scanner. The MR data were acquired with a 16-channel RF head coil, and the human PET data were acquired with an average dose 233MBq of [18-F] FDG 
infused into the subject over the course of the scan at a rate of 36mL/hr \cite{jamadar2020}. The raw data for a particular time image was rebinned into sinograms, with 157 angular projections and 219 radial bins, for PET image reconstruction; and the corresponding MR image was used as training data for constructing the kernel and Laplacian matrices and dictionary learning.


To reconstruct PET images with better quality, we used manufacturer software to evaluate correction factors for random scatters, attenuation and detector normalization, and used these in the reconstruction process of all methods.


Based on the parameters from simulation studies, the parameters of all four methods were individually fine-tuned to achieve their respective best visual performance. The images thus reconstructed are shown in the upper-left panes of Fig. \ref{fig10} and Fig. \ref{fig11}. Compared with MLEM+F and KEM, KREM and MKREM images have better visual effect, with clearer edges and details, and MKREM also further outperforms KREM.

To validate the robustness of KREM and MKREM, we detuned the best iteration numbers and the best neighborhood sizes (excluding MLEM+F) of all four methods. The reconstructed images of the four methods under three different scenarios of the detuned parameters are shown in the upper-right, lower-left and lower-right panes of Fig. \ref{fig10} and Fig. \ref{fig11}. It can be seen that the MLEM+F and KEM images have changed quite noticeably under the detuned parameters, becoming noisy and showing image artifacts. In contrast, the changes in KREM and MKREM images, especially in MKREM images, are barely noticeable. These demonstrate the superior robustness of KREM and MKREM and the sensitivity of MLEM+F KEM to their tuning parameters in real life applications.

The above tests were conducted using Matlab 2016b on a PC with an Intel Core i7 3.60GHz CPU. The computation time with the uncompiled Matlab code was 34 s for KREM and 128 s for MKREM over 40 iterations.

\section{Discussions}
\label{discus}

The results of simulation studies and in vivo data tests clearly show that KREM and MKREM outperform MLEM+F and KEM, achieving lower AMSE, faster convergence speed, simultaneously lower bias and variance, lower sensitivity to the neighborhood size $J_a$, and high robustness to the iteration number. Moreover, all these advantages are more pronounced for MKREM.

The superior reconstruction performance of KREM and MKREM over MLEM+F and KEM can be attributed to the anatomical information of training images embedded in and enforced by the kernel space dictionary and the graph Laplacian quadratic in the two regularizers of KREM and MKREM. The KREM and MKREM's much faster convergence speed and high robustness to the iteration number can be attributed to their two regularizers. And the better performance of MKREM over KREM can be attributed to its distinct multi-kernel matrices used in the dictionary learning, Laplacian matrix construction and MKREM algorithm (\ref{Imgrecon}).

For the low count case in Fig. \ref{fig8}(a), the larger the $J_a$, the lower the MSE. This is because denoising is the key to reduce AMMSE in low count case. With larger $J_a$, the algorithms use more neighboring pixels to reconstruct a pixel, which smooths out the noise and hence decreases AMMSE. While for the high count case in Fig. \ref{fig8}(b), the algorithms achieve minimum AMMSE at a smaller $J_a$, eg $J_a = 21$ for MKREM, a further increase of $J_a$ will increase AMMSE. This is because the key to reduce AMMSE in this case is not denoising but edges preserving. A larger $J_a$ could blur the details and edges and thereby increases AMMSE. Moreover, larger $J_a$ leads to denser kernel matrix and increases greatly computation cost. Whereas, $J_b < J_a$ is generally needed in dictionary learning for the kernel $\bm K_{Mb}$ to capture the image details in high resolution image priors, eg $J_b =3$ in our simulation studies and in vivo data tests.

\section{Conclusion}
\label{conclu}

This paper has presented a novel solution to the problems of the recently emerged kernelized MLEM methods in potentially large reconstruction error and high sensitivity to iteration number. The solution has provided a general framework for developing the regularized kernelized MLEM with multiple kernel matrices and multiple kernel space regularizers for different PET imaging applications. The presented general class of multi-kernel matrices and two specific kernel space regularizers have resulted in the KREM and MKREM algorithms. Extensive test and comparison results have shown the superior performance of these new algorithms and their advantages over the kernelized MLEM and other conventional iterative methods in PET image reconstruction.

The multiple kernel and multiple regularizer framework of the MKREM algorithm (\ref{KRMAP}) has offered many possibilities for further performance improvements. For example, instead of the MR anatomical image priors, we can also use other prior data, alone or mixed with MRI data, to construct the multi-kernel matrices, kernel image dictionary and Laplacian matrix for (\ref{KRMAP}). We have mixed both the spatial and temporary priors to construct these matrices and have obtained excellent results in dynamic PET imaging, with much improved spatial and temporal resolutions. Moreover, we may use the multiple kernel leaning approach developed in machine learning \cite{onen2011,cortes2009} to learn more general kernel polynomials from prior data and hence obtain the optimal multi-kernel matrices for an application, instead of constructing them by careful selection of the kernels and their combinations manually. These results will be reported in future works.

\end{document}